\documentclass[aps,prb,amssymb,twocolumn,showpacs]{revtex4-1}

\newcommand{\bk}{{\bf k}}

\newcommand{\bq}{{\bf q}}

\newcommand{\bn}{{\bf n}}

\newcommand{\br}{{\bf r}}

\newcommand{\ka}{k_{\alpha}}
\newcommand{\kb}{k_{\beta}}
\newcommand{\kd}{k_{\delta}}
\newcommand{\kg}{k_{\gamma}}

\newcommand{\qd}{q_{\delta}}
\newcommand{\qg}{q_{\gamma}}

\usepackage{graphics,graphicx,amsmath}

\usepackage{tabularx,float}

\usepackage{subfigure}

\usepackage{bm}
\usepackage{wasysym}

\usepackage{bm}
\usepackage{color}
\usepackage{verbatim}

\begin{document}

\bibliographystyle{apsrev4-1}

\date{\today}

\author{R. Rold\'{a}n, A. Fasolino, K. V. Zakharchenko and M. I. Katsnelson}

\affiliation{\centerline{Institute for Molecules and Materials, Radboud University Nijmegen, Heyendaalseweg 135, 6525 AJ Nijmegen, The Netherlands}}

\title{Suppression of anharmonicities in crystalline membranes by external strain}

\begin{abstract}
In practice, physical membranes are exposed to a certain amount of external strain (tension or compression), due to the environment where they are placed. As a result, the behavior of the phonon modes of the membrane is modified. We show that anharmonic effects in stiff two-dimensional membranes are highly suppressed under the application of tension. For this, we consider the anharmonic coupling between bending and stretching modes in the self-consistent screening approximation (SCSA), and compare the obtained height-height correlation function in the SCSA to the corresponding harmonic propagator. The elasticity theory results are compared to atomistic Monte Carlo simulations for a graphene membrane under tension. We find that, while rather high values of strain are needed to avoid anharmonicity in soft membranes, strain fields less than 1\% are enough to suppress all the anharmonic effects in stiff membranes, as graphene.

\end{abstract}


\maketitle

\section{Introduction}

The study of mechanical properties of physical membranes, which are two-dimensional surfaces embedded in three-dimensional space, has a prominent experimental platform in graphene,\cite{NF04} a single layer of carbon atoms arranged in a hexagonal crystalline order.\cite{CG09,VKG10} Graphene is a stiff membrane, which presents long wavelength modulations of the out-of-plane displacements, commonly referred as ripples.\cite{MR07,BG08,FLK07} The impact of corrugation on electronic transport, as well as the mechanical properties of graphene, are subjects of intense investigation.\cite{KC08,MO08,G09b,EK10,SGG11,MO10,CG10} 

Of special interest is to understand the effects of an external strain applied to the membrane.\cite{GP88} This is so because most of the graphene samples are subject to some finite amount of strain, due either to the pinning to the substrate (for samples on SiO$_2$, for example) or to the electrostatic force due to the gate on suspended samples. In particular, the existence of tension affects the dispersion of flexural phonons.\cite{MO10,CG10} In fact, whereas in the harmonic approximation and in the absence of strain, the dispersion relation of flexural phonons is quadratic, $\omega_{fl}(q)\sim q^2$, strain introduces a characteristic wave-vector $q_s$ where the dispersion changes from linear (for $q<q_s$) to quadratic (for $q>q_s$). However, anharmonic coupling between bending and stretching modes is important and leads to a further renormalization of the mode dispersion, especially at long wavelengths.\cite{NP87,AL88,DR92,KM09,LF09,BH10}

In this paper, we study the effect of tension on the flexural phonons of a 2D membrane. For this aim, we include a strain field in the free energy, which is studied first in the harmonic approximation and then including anharmonic effects, using the self-consistent screening approximation\cite{DR92,G09c,ZRFK10} (SCSA). The results of a stiff membrane as graphene are compared to those for a softer membrane, for which the bending rigidity has been highly reduced. The validity of the continuum elastic theory is checked by comparing the SCSA results to atomistic Monte Carlo (MC) simulations. Our numerical results show that, for stiff membranes as graphene, small amounts of tension can be used to suppress the anharmonic effects. The case of compressional strain is much more complex due to its highly non-linear behavior,\cite{SS02,CM03,MG99,BD11} and it cannot be accounted for in the SCSA. However, we can still use atomistic MC simulations for this case and, in fact, we find a highly non-trivial behavior for the correlation function of a compressed graphene membrane, with no crossover to a power-law behavior as in the tensioned case.

The paper is organized as follows. In Sec. \ref{Sec:Harmonic} we discuss, in the harmonic approximation, the effect of an external strain in the system, and compare the results to the unstrained case. In Sec. \ref{Sec:SCSA} we consider the anharmonic coupling between bending and stretching modes in the SCSA. In Sec. \ref{Sec:MC} the results of the continuum elastic theory are compared to atomistic MC simulations for the height-height correlation function. The MC simulation for a compressed membrane is also included. The main conclusions of our work are summarized in Sec. \ref{Sec:Conclusions}.

\section{Harmonic approximation}\label{Sec:Harmonic}

In the absence of any external strain, the flat phase of a 2D membrane at sufficiently long scales is well described by a free energy that is a sum of a bending and a stretching part\cite{NPW04} 
\begin{equation}\label{Eq:F_u=0}
{\cal F}[{\bf u},h]=\frac{1}{2}\int d^2\br \left[\kappa \left(\nabla^2h\right)^2+2\mu u_{\alpha\beta}^2+\lambda u_{\alpha\alpha}^2\right]
\end{equation}
where $\kappa$ is the bending rigidity, $\lambda$ and $\mu$ are the first Lam\'e constant and the shear modulus, respectively,\footnote{In most part of this paper, we use the typical parameters for graphene at room temperature: $\kappa\approx1.1$eV, $\lambda\approx2.4$eV\AA$^{-2}$ and $\mu\approx9.95$eV\AA$^{-2}$.} and $u_{\alpha\beta}$ is the {\it internal} strain tensor
\begin{equation}\label{Eq:StrainTensor}
u_{\alpha\beta}\approx \frac{1}{2}(\partial_{\alpha}u_{\beta}+\partial_{\beta}u_{\alpha}+\partial_{\alpha}h\partial_{\beta}h).
\end{equation} 
In the harmonic approximation, the bending and stretching modes are decoupled. From Eq. (\ref{Eq:F_u=0}), one can calculate the correlation function for the out-of-plane displacements $h(\br)$ which, in Fourier space, reads
\begin{equation}\label{Eq:G0_u=0}
\langle |h(\bq) |^2 \rangle_{u=0}=\frac{k_BT}{\kappa q^4},
\end{equation}
where $k_B$ is the Boltzman constant, $T$ is the temperature, and the suffix $u=0$ in the average denotes the absence of any external strain. The effect of an external strain applied to the membrane is modeled by the inclusion of a new term in the theory, $\tau_{\alpha\beta}$, that couples to the internal strain tensor. Therefore, we use the following expansion for the free energy
\begin{equation}\label{Eq:F}
{\cal F}[{\bf u},h,\tau_{\alpha\beta}]=\frac{1}{2}\int d^2\br \left[\kappa \left(\nabla^2h\right)^2+2\mu u_{\alpha\beta}^2+\lambda u_{\alpha\alpha}^2+\tau_{\alpha\beta}u_{\alpha\beta}\right]
\end{equation}
where $\tau_{\alpha\beta}=\lambda\delta_{\alpha\beta}u^{ext}_{\alpha\beta}+2\mu u^{ext}_{\alpha\beta}$ is expressed in terms of the {\it external} strain tensor $u^{ext}_{\alpha\beta}$.  In the harmonic approximation, the Fourier component of the height-height correlation function reads simply
\begin{equation}\label{Eq:G0}
G_0(\bq)\equiv \langle |h(\bq) |^2 \rangle_{u}=\frac{k_BT}{q^2\left (\kappa q^2+\lambda u^{ext}_{\alpha\alpha}+2\mu u^{ext}_{\alpha\beta}\frac{q_{\alpha}q_{\beta}}{|\bq|^2}\right)}.
\end{equation}
For an isotropic expansion and in the long wavelength limit, we can approximate 
\begin{equation}
u^{ext}_{\alpha\beta}=u\delta_{\alpha\beta}
\end{equation}
where $u=\delta S/2S$ accounts for the uniform dilation of the membrane, where $S$ is the membrane surface and $\delta S$ is the change in area due to the application of strain. This reduces $G_0(\bq)$ to
\begin{equation}\label{Eq:G0uniform}
G_0(\bq)=\frac{k_BT}{q^2[\kappa q^2+2(\lambda+\mu)u]}
\end{equation}
 where $\tau=2(\lambda+\mu)u$ is the stress of the system. Notice that for the unstrained case ($u=0$), as given by Eq. (\ref{Eq:G0_u=0}), the mean square amplitude of the out-of-plane displacement diverges, in the harmonic approximation, as\cite{NPW04} $\langle h^2\rangle_{u=0} \propto L^2$, where $\langle h^2\rangle=\sum_{\bq}\langle | h(\bq)|^2\rangle$ and $L$ is the sample size. Furthermore, the normal-normal correlation $\langle \bn(\br)\cdot\bn(0)\rangle$ diverges logarithmically as $r\rightarrow \infty$. However, for a membrane under uniform dilation, we can use Eq. (\ref{Eq:G0uniform}) and obtain
\begin{equation}
\langle h^2\rangle= \frac{k_BT}{8\pi(\lambda+\mu)u}\log\left(1+\frac{2(\lambda+\mu)u}{\kappa q_{min}^2}\right)\propto \frac{\log (L^2u)}{u}
\end{equation}
where $q_{min}=2\pi/L$ is an infrared cutoff of the order of the inverse sample size $L$. Furthermore, from the normal-normal correlation function $\langle |{\bf n}(\bq)|^2\rangle=k_BT/[\kappa(q^2+q_u^2)]$, where $q_u=\sqrt{2(\lambda+\mu)u/\kappa}$ we obtain that 
\begin{equation}
\langle \bn(\br)\cdot\bn(0)\rangle = \frac{k_BT}{\kappa}\int \frac{d^2\bq}{(2\pi)^2} \frac{e^{i\bq\cdot\br}}{q^2+q_u^2}=\frac{ k_BT}{2\pi\kappa}K_0(q_ur),
\end{equation}
where $K_0(x)$ is a modified Bessel function of the second kind. Taking into account that, for $x\gg 1$, $K_0(x)\approx \sqrt{\pi/2x}e^{-x}$, we obtain that, for $r\rightarrow \infty$
\begin{equation}
\langle \bn(\br)\cdot\bn(0)\rangle \approx \frac{ k_BT}{\kappa}\frac{e^{-q_ur}}{\sqrt{8\pi q_ur}}.
\end{equation}
Therefore, the application of external strain, as expected, guarantees the long range 2D order of the membrane. Furthermore, $q_u^{-1}$ defines a length-scale that separates the strain dominated from the unstrained regions of the correlation functions. 
\newline

\section{Anharmonic effects: continuum elastic theory in the SCSA}\label{Sec:SCSA}

\begin{figure}[t]
\centering
\includegraphics[width=0.47\textwidth]{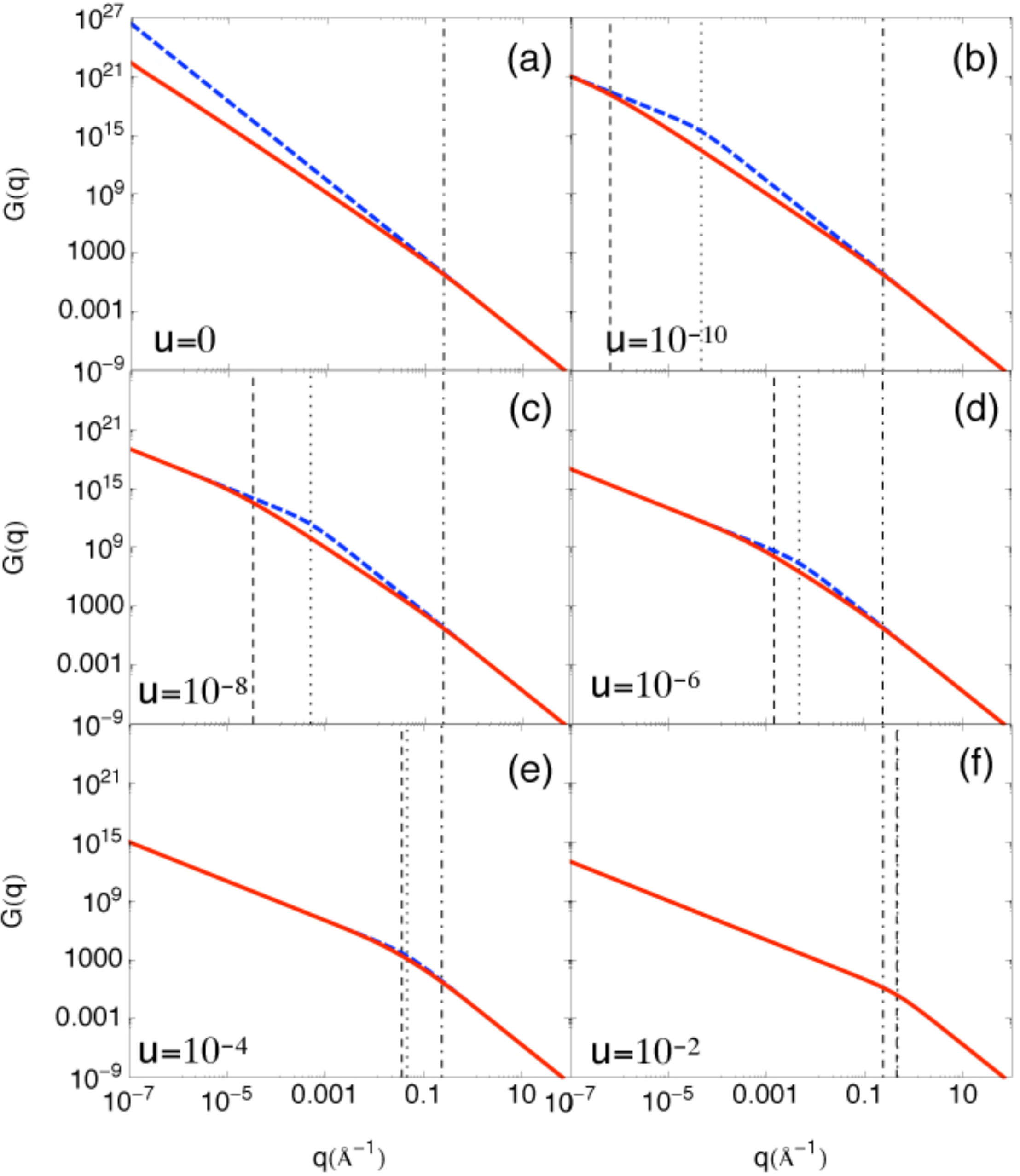}
 \caption{(Color online) Correlation function in the harmonic approximation, $G_0(\bq)$ (blue dashed line) and renormalized correlation function in the SCSA $G(\bq)$ (full red line) for several values of strain $u$ from 0 to $10^{-2}$. Dotted-dashed vertical lines indicate $q_c\approx 0.24$ \AA$^{-1}$, according to the Ginzburg criterion, Eq. (\ref{Eq:GC}). Dotted vertical lines indicates the position of $q_s^h$ and dashed vertical lines point the position of $q_s$ (see text).}
 \label{Fig:G-u}
\end{figure}

In the previous section we have seen that the application of an external strain stabilizes, even in the harmonic approximation, the flat phase of a 2D membrane. But even in the absence of strain, it is known that the flat phase is stable. This is due to the anharmonic coupling between bending and stretching modes.\cite{NPW04} Therefore, anharmonic effects lead to a further renormalization of the characteristic lengths and elastic constants discussed in the previous section. In the following, we study the effect of anharmonicity in the correlation function of a strained 2D membrane. First, one notices that the in-plane phonons in the free energy Eq. (\ref{Eq:F}) can be integrated out exactly, what allows us to write an effective action in terms only of the $h$ fields\cite{RN91}
\begin{eqnarray}\label{Eq:Freal}
{\cal F}_{eff}[h,\tau_{\alpha\beta}]&=&\int d^2\br  \left[\frac{1}{2}\kappa\left( \nabla^2h \right)^2 + \tau_{\alpha\beta}\partial_{\alpha}h\partial_{\beta}h\right ] \nonumber\\
&+&\frac{1}{8}Y\int d^2\br  (P_{\alpha\beta}^T\partial_{\alpha}h\partial_{\beta}h+P_{\alpha\beta}^Tu^{ext}_{\alpha\beta})^2 \nonumber\\
\end{eqnarray}
where $Y=4\mu(\mu+\lambda)/(2\mu+\lambda)$ is the 2D Young modulus and $P_{\alpha\beta}^T=\delta_{\alpha\beta}-\partial_{\alpha}\partial_{\beta}/\nabla^2$ is the transverse projection operator. Eq. (\ref{Eq:Freal}) can be expressed in terms of the Fourier components of the height field, $h(\bq)$. To the lowest order in $u^{ext}$ we obtain
\begin{widetext}
\begin{eqnarray}\label{Eq:FFourier}
{\cal F}_{eff}[h,\tau_{\alpha\beta}]&=&\frac{1}{2}\int \frac{d^2\bk}{(2\pi)^2}  k^2\left(\kappa k^2+\lambda u_{\alpha\alpha}^{ext}+2\mu \frac{k_{\alpha}k_{\beta}}{k^2}u_{\alpha\beta}^{ext}\right) |h(\bk)|^2\nonumber\\
&+&\frac{1}{8}Y\int \frac{d^2\bk_1}{(2\pi)^2}\int \frac{d^2\bk_2}{(2\pi)^2}\int \frac{d^2\bk_3}{(2\pi)^2} P_{\alpha\beta}^TP_{\gamma\delta}^T(\bq) k_{1\alpha}k_{2\beta}k_{3\gamma}k_{4\delta}[h(\bk_1)h(\bk_2)][h(\bk_3)h(\bk_4)]\nonumber\\
&+&\frac{1}{4}Y\int \frac{d^2\bk_1}{(2\pi)^2}\int \frac{d^2\bk_2}{(2\pi)^2} P_{\alpha\beta}^TP_{\gamma\delta}^T(\bq)k_{1\alpha}k_{2\beta}u^{ext}_{\gamma\delta}(\bq)h(\bk_1)h(\bk_2)
\end{eqnarray}
\end{widetext}
where $\bq=\bk_1+\bk_2$ and in the second term $\bk_1+\bk_2+\bk_3+\bk_4=0$. The first line of Eq. (\ref{Eq:FFourier}) is nothing but the bending part of the free energy in the harmonic approximation, from which we have defined the non-interacting correlation function $G_0(\bq)$, as in Eq. (\ref{Eq:G0}) of the previous section. The second and third lines build the interaction term of the theory. The second line of Eq. (\ref{Eq:FFourier}) accounts for the four point vertex, whereas the last term of this equation leads to a two-point vertex that renormalize the propagator. This problem is similar to that of a polymerized membrane with long-range disorder, and can be treated in the SCSA.\cite{DR93}

 For stiff membranes as graphene, the anharmonic effects are quickly suppressed under the application of strain. Therefore, we assume that the renormalization of the propagator due to the vertices associated to the last term of Eq. (\ref{Eq:FFourier}) is weak, and that we can neglect this class of diagrams in our calculations. We will see in Sec. \ref{Sec:MC} that the correlation functions calculated with this assumption in the SCSA agree well with those obtained from atomistic MC simulations, justifying the simplification. Then, the renormalized correlation function can be calculated from a closed self-consistent set of two coupled integral equations for the self-energy\cite{DR92,G09c} 
\begin{eqnarray}\label{Eq:SCSA}
\Sigma(\bk)&=&2\ka\kb\kg\kd\int \frac{d^2\bq}{(2\pi)^2} {\tilde R}_{\alpha\beta,\gamma\delta}(\bq)G(\bk-\bq)\label{Eq:Sigma}\\
{\tilde R}_{\alpha\beta,\gamma\delta}(\bq)&=&R_{\alpha\beta,\gamma\delta}(\bq)-R_{\alpha\beta,\mu\nu}(\bq)\Pi_{\mu\nu,\mu'\nu'}(\bq){\tilde R}_{\mu'\nu',\gamma\delta}(\bq)\nonumber\\
\label{Eq:Vertex}
\end{eqnarray}
where $G^{-1}(\bq)=G_0^{-1}(\bq)+\Sigma(\bq)$ is the inverse of the {\it dressed} propagator, $\Pi_{\alpha\beta,\gamma\delta}(\bq)$ are the vacuum polarization functions, 
\begin{equation}
\Pi_{\alpha\beta,\gamma\delta}(\bq)=\int \frac{d^2 \bk}{(2\pi)^2} \ka\kb(\kg-\qg)(\kd-\qd) G(\bk)G(\bq-\bk),
\end{equation}
$R_{\alpha\beta,\gamma\delta}(\bq)=(Y/2-\mu)P_{\alpha\beta}^TP_{\gamma\delta}^T+(\mu/2)(P_{\alpha\gamma}^TP_{\beta\delta}^T+P_{\alpha\delta}^TP_{\beta\gamma}^T)$ is the unrenormalized four-point interaction vertex and ${\tilde R}_{\alpha\beta,\gamma\delta}(\bq)$ is the screened interaction.

 The set of equations (\ref{Eq:Sigma})-(\ref{Eq:Vertex}) can be solved, at any wave-vector, following the method introduced in Ref. \onlinecite{ZRFK10}.  Fig. \ref{Fig:G-u} shows the momentum dependence of $G_0(\bq)$ and $G(\bq)$ for different values of strain, $u=0,...,10^{-2}$. First, one notices in Fig. \ref{Fig:G-u}(a) that, without strain, the harmonic approximation is valid only in the short wavelength region, where $G(\bq)\approx G_0(\bq)$.\cite{ZRFK10} The Ginzburg criterion, which considers only the first order correction to the correlation function, allows to estimate the characteristic wave-vector $q_c$ above which the harmonic behavior applies. For 2D membranes, $q_c$ is approximately given by\cite{NPW04}
\begin{equation}\label{Eq:GC}
q_c=\sqrt{\frac{3k_BTY}{8\pi\kappa^2}}.
\end{equation}
For the parameters of graphene, $q_c\approx 0.24~\rm\AA^{-1}$ at room temperature, a value which is shown by the vertical dot-dashed lines in Fig. \ref{Fig:G-u}. This is the characteristic wave-vector at which the renormalized correlation function $G(\bq)$ (full red line) separates from the harmonic approximation $G_0(\bq)$ (dashed blue line), pointing out that anharmonic effects are very important at long length scales. If we consider the effect of external strain on the membrane, we still can distinguish between the harmonic and the anharmonic regimes, as shown in Fig. \ref{Fig:G-u}(b)-(f).\footnote{The inclusion of the last term of Eq. (\ref{Eq:FFourier}) in the calculation would lead to an external strain dependence of $q_c$, effect that is neglected in the simple approximation used to obtain Eq. (\ref{Eq:GC}).} However, there also exists a characteristic scale at which the behavior of flexural phonons is dominated by strain effects. This scale manifests itself as a change in the slope of the correlation functions: $G_0(\bq)$ changes from $\sim q^{-4}$ to $q^{-2}$ and $G(\bq)$ changes from $q^{-4+\eta}$ also to $q^{-2}$, where $\eta$ is a characteristic exponent. In fact, we can observe from Fig. \ref{Fig:G-u}(b)-(f) how the region of intermediate momenta where the two functions $G_0(q)$ and $G(q)$ are different, and therefore anharmonic effects are important, is reduced as the value of $u$ grows. From these results we see that very small amounts of strain are enough to suppress the anharmonic effects.

\begin{figure}[t]
\centering
\includegraphics[width=0.40\textwidth]{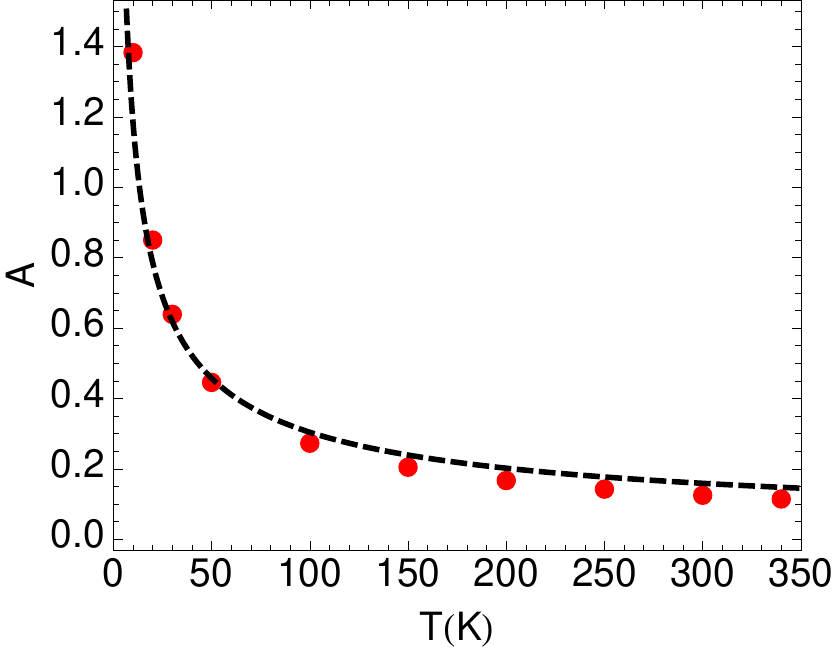}
 \caption{The temperature dependence of the parameter $A$ of Eq. (\ref{Eq:BR}), obtained from the SCSA correlation functions at different temperatures (red dots). The dashed line is a fitting to Eq. (\ref{Eq:A}). For these plots, we have used the wave-vector $q=10^{-2}\AA^{-1}$.}
 \label{Fig:A-T}
\end{figure}

These results can be used to study the effect of strain on flexural (out-of-plane) phonons. The dispersion relation for flexural phonons of a 2D membrane under isotropic tension can be written as
\begin{equation}
\omega_{fl}(\bq)=\sqrt{\frac{\kappa(\bq)}{\rho}q^4+u\frac{2(\lambda+\mu)}{\rho}q^2}
\end{equation}
where $\rho$ is the density and $\kappa(\bq)$ is the bending rigidity. In the harmonic approximation, $\kappa(\bq)\equiv\kappa$ and the dispersion changes from linear to quadratic at a wave-vector equal to 
\begin{equation}
q_s^h=\sqrt{\frac{2u(\mu+\lambda)}{\kappa}}.
\end{equation}
This characteristic wave-vector is denoted by the vertical dotted lines in Fig. \ref{Fig:G-u}(b)-(f).\footnote{Notice that $q_s^h$ coincides with the wave-vector $q_u$ discussed in the previous section.} However, anharmonic effects are important at long scales. To obtain analytical results, we use the effective Dyson equation for the correlation function\cite{FLK07,ZRFK10}
 \begin{equation}\label{Eq:Ga}
 G_a^{-1}(\bq)=G_0^{-1}(\bq)+\Sigma_a(\bq),
 \end{equation} 
 where $G_a(\bq)$ is an approximated correlation function dressed by the self-energy $\Sigma_a(\bq)$, which is approximated by
\begin{equation}\label{Eq:Sigmaa}
 \Sigma_a(\bq)=Aq^4\left(\frac{q_0}{q} \right)^{\eta},
 \end{equation}
where $A$ is some numerical factor, $\eta\approx 0.82$,\cite{DR92} and $q_0=2\pi\sqrt{Y/\kappa}$. From the approximation Eq. (\ref{Eq:Ga}) one can obtain the renormalized bending rigidity
 \begin{equation}\label{Eq:BR}
 \kappa_R(\bq)=\kappa+k_BTA\left(\frac{q_0}{q}\right)^{\eta}.
 \end{equation} 
It is important to mention that the coefficient $A$ is temperature dependent. Notice that anharmonic effects are present in $G_a(\bq)$ below a characteristic wave-vector $q^*$, which is solution of $\Sigma_a(q^*)\approx G_0^{-1}(q^*)$. Assuming that $q_c$ is the only crossover wave-vector from harmonic to anharmonic behavior, and that $q^*\simeq q_c$, then one can easily obtain that the temperature dependence of the parameter $A$ in Eq. (\ref{Eq:Sigmaa}) follows the power-law $A\propto (k_BT/\kappa)^{\frac{\eta}{2}-1}$.\cite{K10} By fitting the SCSA correlation function for different temperatures to Eq. (\ref{Eq:Ga}), we find the dependence of the parameter $A$ on temperature, and the results are shown in Fig. \ref{Fig:A-T} (red dots). This allows to define an approximate expression for the adimensional parameter $A$, which is (using the elastic constants valid for graphene)
\begin{equation}\label{Eq:A}
A\approx 4.6 T[K]^{\frac{\eta}{2}-1}
\end{equation} 
where $T[K]$ is the temperature expressed in Kelvin. The results are shown in Fig. \ref{Fig:A-T} by the dashed line, which fits rather well the values of $A$ obtained numerically. This confirms that the assumption $q^*\simeq q_c$ is indeed valid within the SCSA. The main message is that the bending rigidity Eq. (\ref{Eq:BR}) grows with temperature as 
\begin{equation}\label{Eq:kappaT}
\kappa_R\propto T^{\eta/2}.
\end{equation}
This power-law behavior is similar to the temperature dependence found by Monte Carlo simulations in the harmonic regime.\cite{FLK07,ZLKF10} However we emphasize that here we are assuming that the parameters $\kappa$, $\mu$ and $\lambda$ of the Hamiltonian (\ref{Eq:F_u=0}) are independent of temperature. However, while $\lambda$ and $\mu$ are only weakly dependent on $T$, the temperature dependence of the bending rigidity $\kappa$ found in MC simulations is rather strong,\cite{ZKF09} and this is not accounted for by Eq. (\ref{Eq:kappaT}). The origin of this $T$-dependence is probably beyond the continuum medium approximation, and it lies beyond the scope of this work.

 Then, from Eq. (\ref{Eq:BR}) one observes that the slope of the dispersion relation $\omega_{fl}(\bq)$ changes from $\sim q$ to $\sim q^{2-\eta/2}$ at the wave-vector solution of
\begin{equation}
\left[\kappa+k_BTA\left(\frac{q_0}{q_s} \right)^{\eta}\right]q_s^2=2u(\lambda+\mu).
\end{equation}
The values of $q_s$ for the values of strains studied here are shown by the vertical dashed lines in Fig. \ref{Fig:G-u}(b)-(f). Notice that the characteristic wave-vector obtained by the approximation to the bending rigidity Eq. (\ref{Eq:BR}) agrees well with the exact result of the SCSA equations.

From the previous expressions, and imposing $q_s=q_c$, it is possible to find the critical value for the strain that is enough to suppress the anharmonic effects completely, at any wave-vector, and this is 
\begin{equation}\label{Eq:uc}
u_c=\frac{3k_BT}{4\pi}\frac{\mu}{\kappa(2\mu+\lambda)}.
\end{equation}
For the parameters of graphene at room temperature, this corresponds to $u_c\approx 0.0025$. In fact, notice that $q_s^h$ and $q_s$ already coincide for $u=10^{-2}>u_c$ and that both are to the right of $q_c$ ($q_s,q_s^h>q_c$) [Fig. \ref{Fig:G-u}(f)], pointing out that anharmonic effects are already absent for this value ($\sim 1\%$) of external strain.

\begin{figure}[t]
\centering
\includegraphics[width=0.47\textwidth]{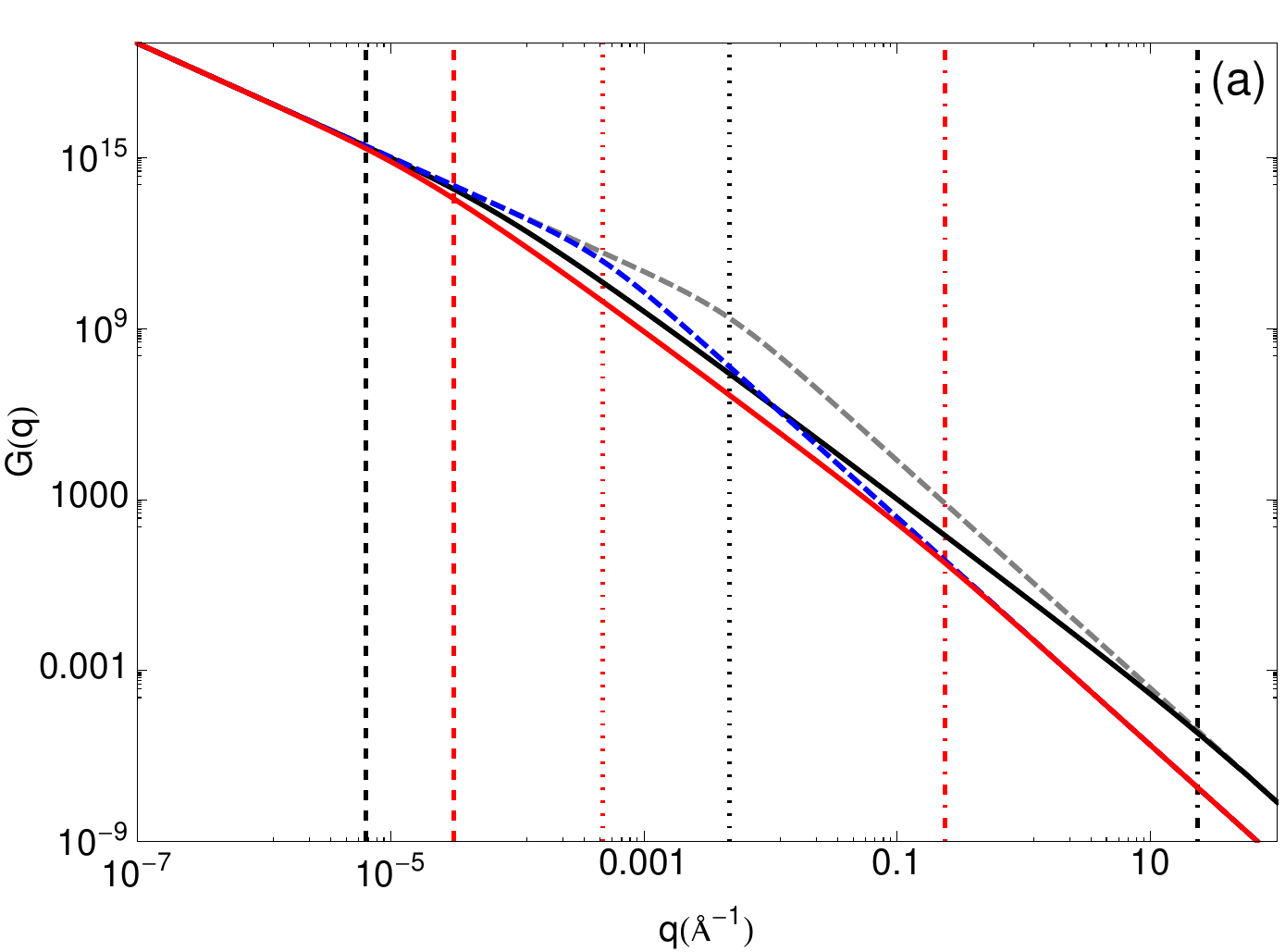}
\includegraphics[width=0.47\textwidth]{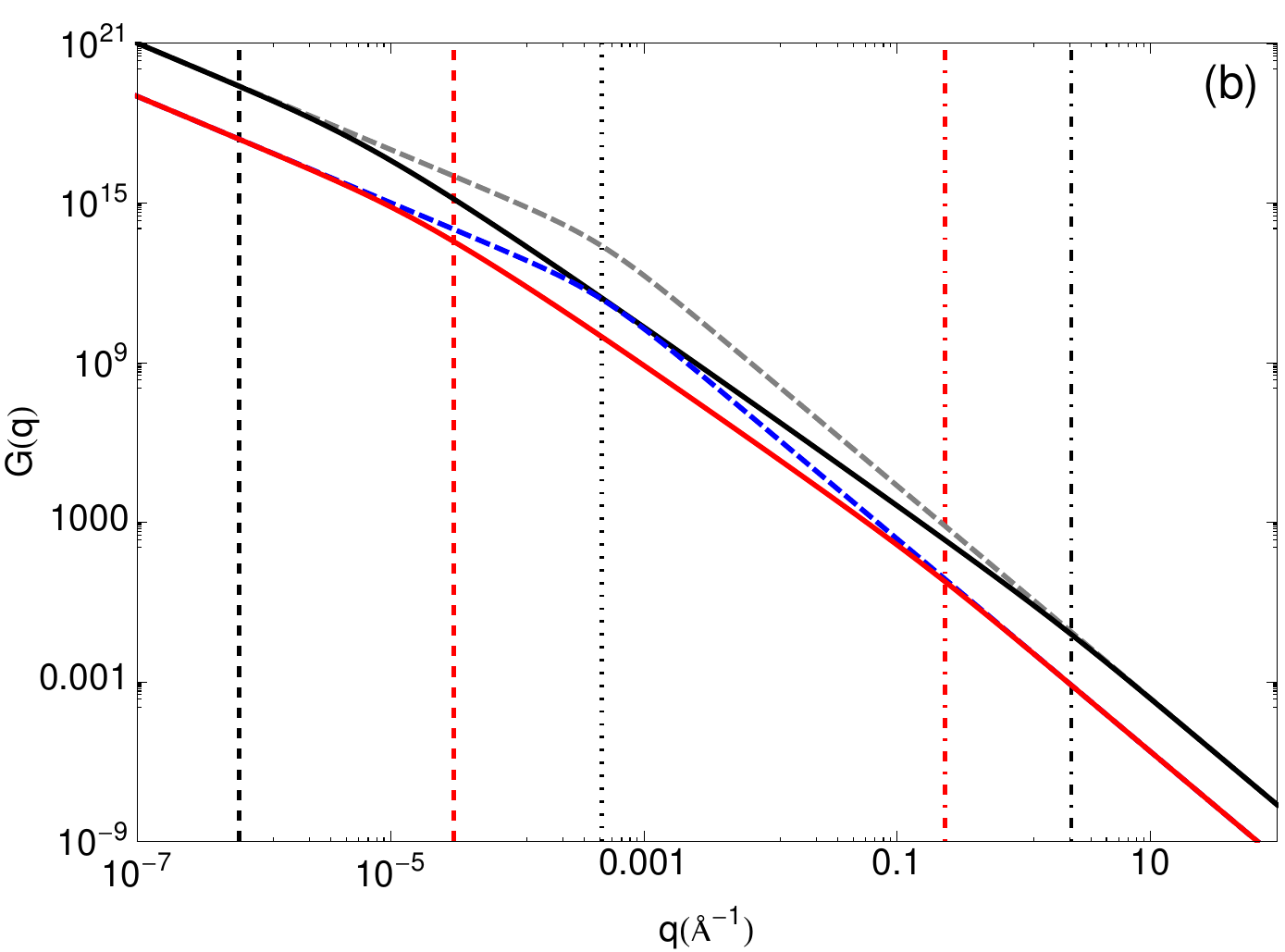}
 \caption{(Color online) $G_0(\bq)$ for graphene (dashed blue line) and for a softer membrane (dashed gray line), and $G(\bq)$ for graphene (full red line) and for a softer membrane (full black line). In the two cases, $u=10^{-8}$. In (a) we have used, for the soft membrane, 1/100 times the bending rigidity $\kappa$ valid for graphene at this temperature, whereas $\mu$ and $\lambda$ are the same as in graphene. In this case, $q_c\approx 0.24~\rm\AA^{-1}$ for graphene (vertical red dotted-dashed line) and $q_c\approx 24~\rm\AA^{-1}$ for the soft membrane (vertical black dotted-dashed line). In (b) we compare the correlation functions for graphene to a softer membrane for which all the elastic constants are reduced to a $1\%$ their value in graphene. In this case, $q_c\approx 2.4~\rm\AA^{-1}$ for the soft membrane, as indicated by the position of the vertical black dotted-dashed line. The vertical dotted and dashed lines represent the positions of $q_s^h$ and $q_s$ respectively, as in Fig. \ref{Fig:G-u}.}
 \label{Fig:graph-soft}
\end{figure}

Finally, we compare the results for graphene to those of a softer membrane. In Fig. \ref{Fig:graph-soft}(a) we show $G_0(\bq)$ and $G(\bq)$ for graphene, and for a membrane with the same $\mu$ and $\lambda$ as graphene, but with a bending rigidity $\kappa$ which is $1\%$ of the corresponding for graphene. In this case, we see that $q_c^{soft}>q_c^{graph}$, as seen by the position of the vertical dotted-dashed lines (red for graphene and black for the soft membrane). This means that anharmonic effects manifest themselves at larger wave-vectors for a soft membrane. Furthermore, the change in slope of the harmonic $G_0(\bq)$ occurs at higher wave-vectors for the soft membrane (vertical black dotted line) as compared to graphene (vertical red dotted line). However, the out-of-plane component of the dispersion for flexural phonons dominates in a wider region of momenta for the soft membrane as compared to graphene, $q_s^{soft}<q_s^{graph}$, as it can be seen by the relative position of $q_s$ for graphene (dashed red line) with respect to that of a soft membrane (dashed black line). Notice that in the latter case, the strain necessary to suppress all the anharmonic effects is $u_c\approx 0.25$, also two orders of magnitude larger than for graphene. In Fig. \ref{Fig:graph-soft}(b) we compare the correlation functions of graphene to those of a softer membrane where not only the bending rigidity $\kappa$, but also the Lam\'e constants $\lambda$ and $\mu$ have been reduced to $1\%$ of their values in graphene. The situation is similar to that described for Fig. \ref{Fig:graph-soft}(a), with the difference that the wave-vectors at which anharmonic and strain effects appear are reduced, as it can be seen by the respective shifts to the left of the black dotted-dashed and dashed lines in Fig. \ref{Fig:graph-soft}(b) with respect to (a). Furthermore, for the parameters of Fig. \ref{Fig:graph-soft}(b), the change in slope of the harmonic $G_0(\bq)$ is the same in the two cases (as shown by the vertical dotted line). Finally, we notice that increasing the temperature acts like an effective softening of the membrane, due to the reduction of the ratio $\kappa/k_BT$.

\section{Comparison to atomistic Monte Carlo simulations}\label{Sec:MC}

\begin{figure}[t]
\centering
\includegraphics[width=0.47\textwidth]{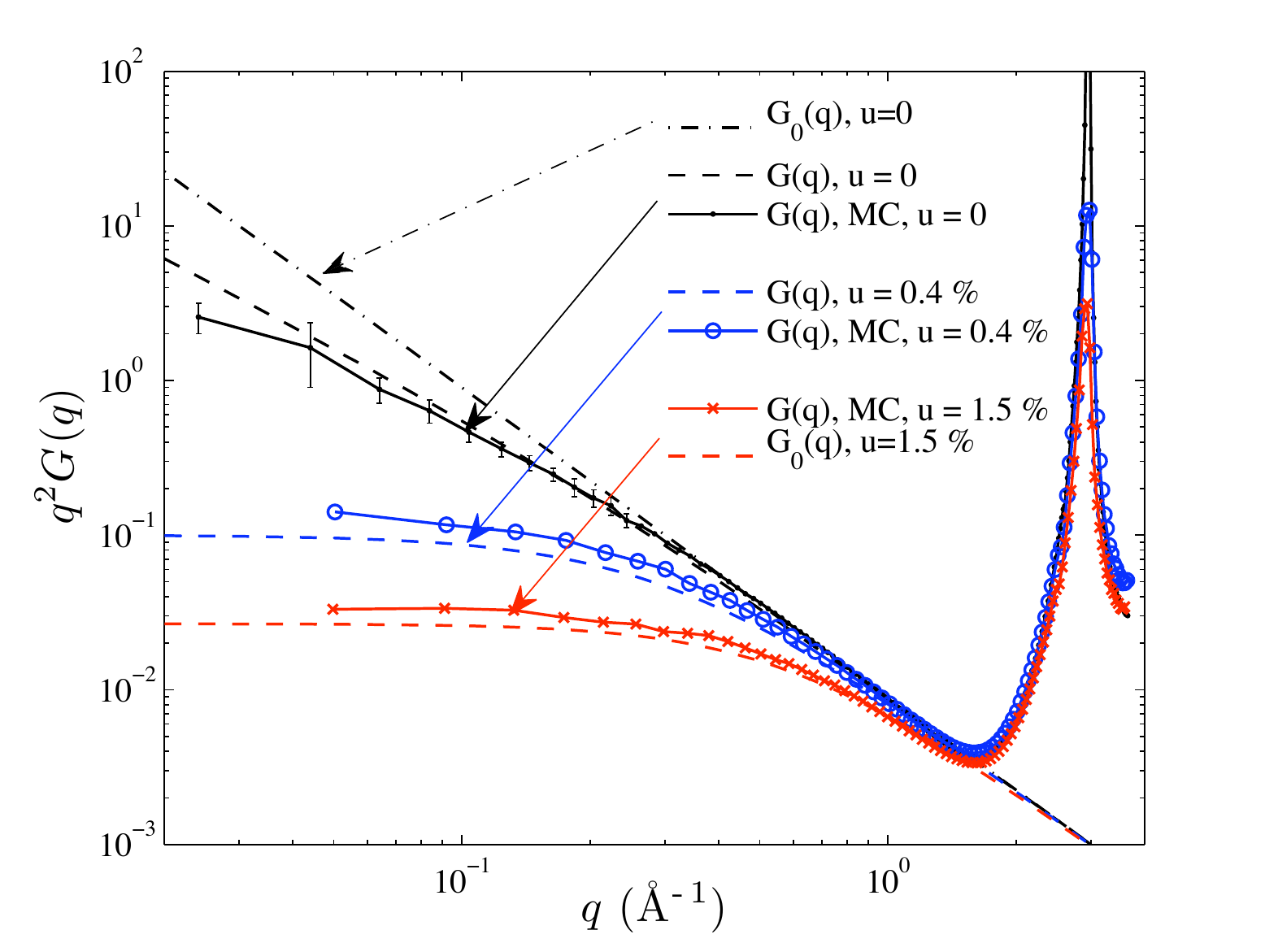}
 \caption{(Color online) Comparison of the normal-normal correlation function $\langle|{\bf n}(\bq)|^2\rangle =q^2G(\bq)$ obtained from continuum elastic theory, as described in Sec. \ref{Sec:Harmonic} and \ref{Sec:SCSA} (dashed lines), to atomistic MC simulations (solid lines), for different values of external strain.}
 \label{Fig:G-MC}
\end{figure}

In this section we compare the results obtained by using the continuum elastic theory methods as described in Sec. \ref{Sec:Harmonic} and \ref{Sec:SCSA}, with the results of the Monte Carlo simulations of graphene. The correlation function $G(\bq)$ for graphene has been calculated as described in Ref. \onlinecite{LF09} for unstrained graphene by means of MC simulations based on an accurate interatomic potential for carbon.\cite{LF05} The simulations are done for a sample of 37888 atoms in a roughly square sample of $314.82 \times 315.24$ \AA$^2$, in the $NPT$ isothermal isobaric ensemble. The simulations for the strained case were done for smaller samples of 8640 atoms ($147.57\times153.36$~\AA$^2$), which limits the range of accessible wave-vectors with respect to the unstrained case. In Fig. \ref{Fig:G-MC} we compare the correlation functions obtained from numerical simulations (full lines) to the SCSA results (dashed lines) for different values of the strain. To highlight the change of slope of $G(\bq)$ due to strain, in Fig. \ref{Fig:G-MC} we plot $q^2G(\bq)$ that becomes flat when $G(\bq)\propto q^{-2}$ as discussed in Sec. \ref{Sec:SCSA}. First, one notices that the $G(\bq)$ calculated by atomistic simulations deviates from those calculated in the continuum limit for wave-vectors close to the Bragg peak at $q = \frac{4\pi}{3a} = 2.94$~\AA$^{-1}$ with $a = 1.42$~\AA~ being the carbon-carbon distance in graphene. We mention also that, for the strained cases, the error bars of the  Monte Carlo simulations are negligible.

Starting from the unstrained case ($u=0$), we see that the MC (full black line) and the SCSA (dashed black line) results agree reasonably well and both deviate from the correlation function in the harmonic approximation (dot-dashed black line) at small wave-vectors, pointing out the importance of anharmonic effects at long scales in unstrained samples.\cite{ZRFK10} For the strained cases, the SCSA and MC results are also comparable, what justify the use of SCSA when dealing with samples under tension. However, we must emphasize that for $0.4\%$ strain, there is almost no difference between $G_0(\bq)$ [as obtained by Eq. (\ref{Eq:G0})] and $G(\bq)$ in the SCSA [full solution of Eq. (\ref{Eq:Sigma})-(\ref{Eq:Vertex})], as discussed in Sec. \ref{Sec:SCSA}, and they are exactly the same for the highest value of strain shown here, $1.5\%$. A more rigorous check of the validity of SCSA would require MC simulations for samples under even weaker strain, which requires several times larger samples to achieve the same accuracy and makes such simulations much more time consuming. Nevertheless, the present results already confirm that rather weak strain is enough to suppress the anharmonicities in stiff membranes as graphene, as it can be seen in Fig. \ref{Fig:G-MC} by the almost flat line-shape of $q^2G(\bq)$ as we move from the Bragg peak towards small wave-vectors of the spectrum, for tensions $\gtrsim 0.4\%$.

\begin{figure}[t]
\centering
\includegraphics[width=0.47\textwidth]{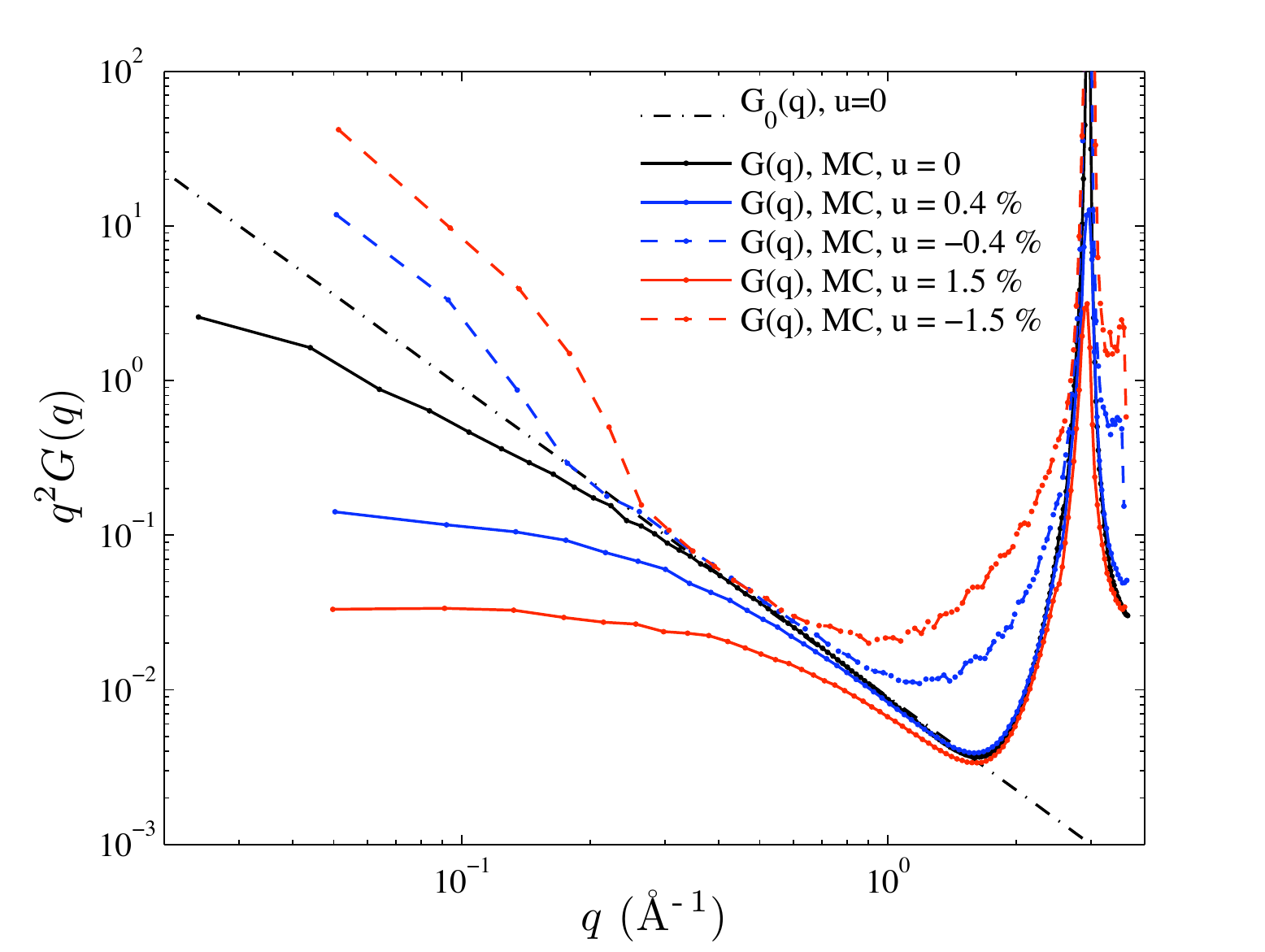}
 \caption{(Color online) Comparison of the normal-normal correlation function $\langle|{\bf n}(\bq)|^2\rangle =q^2G(\bq)$ obtained from atomistic MC simulations for the case of tension (solid lines), and compression (dashed lines), for different values of external strain.}
 \label{Fig:G-MC-Compress}
\end{figure}

The study of a compressed membrane is more delicate because of the fact that its equilibrium state does not correspond to the flat phase any more.\cite{SS02,CM03,MG99,BD11} Therefore, the standard elastic theory that we have used in Sec. \ref{Sec:SCSA} does not apply to this case. However, one can at least study the system by means of Monte Carlo simulations. In Fig. \ref{Fig:G-MC-Compress} we compare the MC results for the correlation function of a tensioned membrane to that of a compressed membrane. There we see that $G(\bq)$ for a compressed membrane does not show the characteristic gradual crossover to another power law as in the tensioned case. Instead, there exist a wave-vector for which the correlation function suffers an abrupt deviation from the harmonic behavior, presenting signatures of a possible first order buckling phase transition. 

\begin{figure*}[t]
\centering
\includegraphics[width=0.77\textwidth]{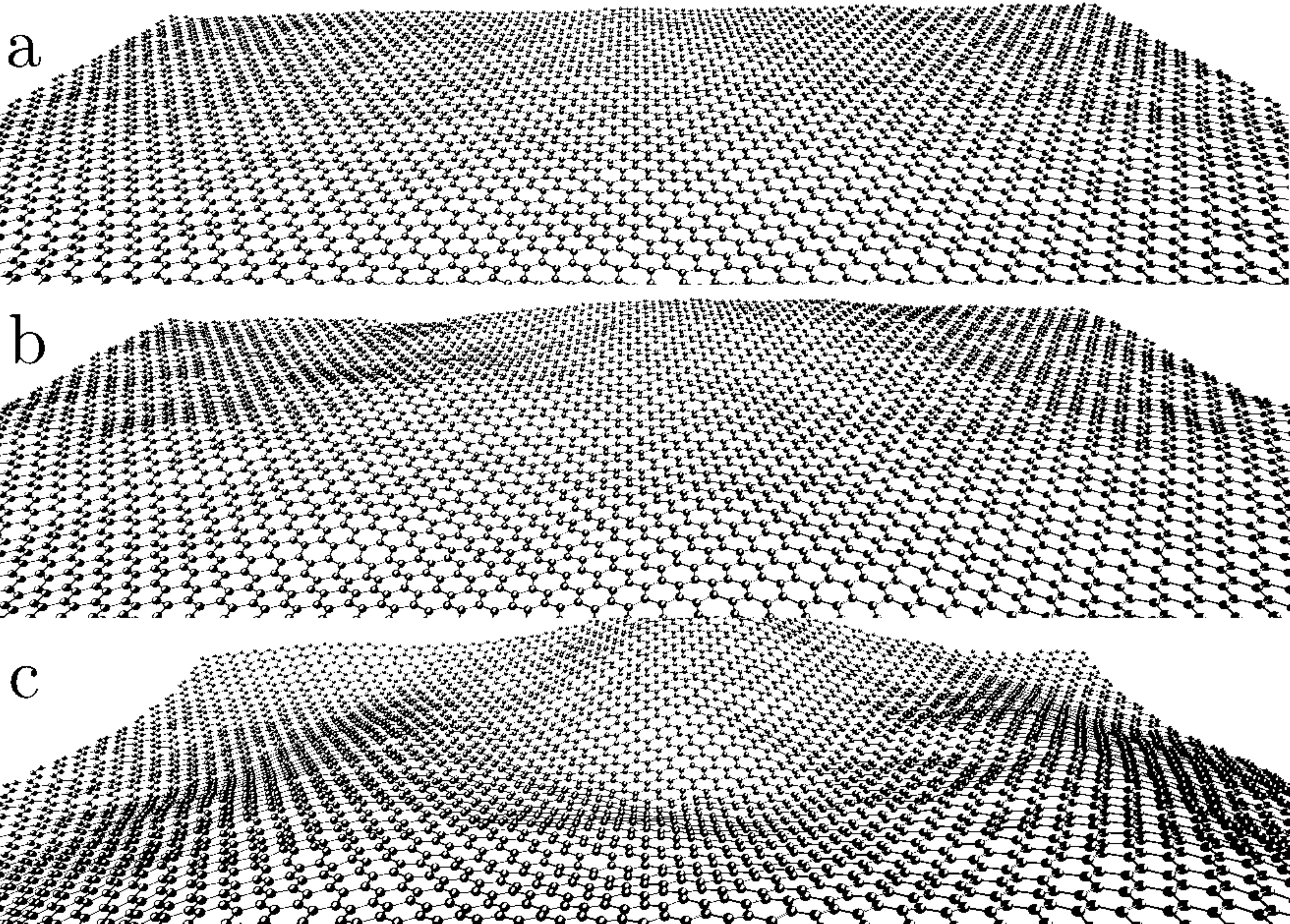}
 \caption{(Color online) Typical Monte Carlo configurations of a graphene sample of 8640 atoms for: a) 1.5\% tension, b) unstrained, and c) 1.5\% compression.}
 \label{Fig:sample}
\end{figure*}

Finally, all our results are graphically summarized in Fig. \ref{Fig:sample}, where we show a snapshot of the Monte Carlo sample for a tensioned graphene membrane [Fig. \ref{Fig:sample}(a)], for an unstrained membrane, Fig. \ref{Fig:sample}(b), and for a compressed graphene sheet, Fig. \ref{Fig:sample}(c). In the first case, the equilibrium state is an almost perfectly flat membrane, for which the anharmonic effects have been suppressed due to the application of tension. The anharmonic coupling between bending and stretching modes is instead important for the case of an unstrained membrane, as the one of Fig. \ref{Fig:sample}(b), which leads to a corrugated low energy phase due to the existence of thermal ripples in the system,\cite{FLK07} and which is well described by means of a continuum elastic theory as the SCSA. However, this theory is not applicable to a compressed membrane as the one shown in Fig. \ref{Fig:sample}(c), for which the sheet buckles into shapes that remove in-plane compression, in order to reduce its elastic energy.\cite{SS02} 

\section{Conclusions}\label{Sec:Conclusions}
In summary, we have studied the effect of external strain in the correlation function of flexural modes in the SCSA. In the presence of strain, three different regimes can be distinguished in the dispersion relation of flexural phonons: $\omega_{fl}(\bq)\sim q$ in the long wavelength limit, $\omega_{fl}(\bq)\sim q^{2-\eta/2}$ in the intermediate range of wave-vectors of the spectrum (where $\eta\approx 0.82$ is a characteristic exponent\cite{DR92}), and finally $\omega_{fl}(\bq)\sim q^2$ at shorter wavelengths. The results show that, for a soft membrane, rather high values of strain are needed to suppress anharmonic effects, whereas for a stiff membrane as graphene, anharmonic effects are completely suppressed by less than 1\% tensile strain. The correlation functions obtained with the SCSA compare well with those calculated from atomistic MC simulations. Taking into account that the scattering of electrons by flexural phonons has been shown to be the main limitation for the charge mobility in suspended graphene,\cite{CG10} our results point that the application of a small tension to the graphene layer would reduce the out-of-plane vibrations that lead to the flexural modes, increasing the mobility of the suspended samples.

 \begin{acknowledgments}
This work is part of the research program of the 'Stichting voor Fundamenteel Onderzoek der Materie (FOM)', which is financially supported by the 'Nederlandse Organisatie voor Wetenschappelijk Onderzoek (NWO)'. We thank the EU-India FP-7 collaboration under MONAMI, and the Netherlands National Computing Facilities foundation (NCF).
 \end{acknowledgments}

\bibliography{BibliogrGrafeno}

\end{document}